\documentclass[11pt,a4paper]{article}
\usepackage[hyperref]{naaclhlt2018}
\usepackage{times}
\usepackage{latexsym}
\usepackage{enumitem}
\setlist{nolistsep}

\usepackage{amsmath}
\usepackage{url}

\usepackage{graphicx}
\usepackage{color}
\usepackage{multirow}
\usepackage{comment}
\usepackage{bm}
\usepackage[font=footnotesize]{caption}
\usepackage{url}

\aclfinalcopy 

\title{Clust-LDA: Joint Model for Text Mining and Author Group Inference}

\author{
	Shaoyang Ning \\
  		{\small Department of Statistics,}\\ {\small Harvard University}  \\
  		{\small \tt shaoyangning@fas.harvard.edu} \\
        \And
  	Xi Qu \\
  		{\small Legendary, Applied Analytics} \\
        {\small \tt squ@legendary.com} \\\And
  	Victor Cai \\
  		{\small Legendary, Applied Analytics} \\
        {\small \tt vc225@cornell.edu} \\\And
  	Nathan Sanders \\ 
  		{\small Legendary, Applied Analytics} \\
        {\small \tt nsanders@legendary.com} \\
  \\}

\date{}

\begin{document}
\maketitle
\begin{abstract}
Social media corpora pose unique challenges and opportunities, including typically short document lengths and rich meta-data such as author characteristics and relationships. This creates great potential for systematic analysis of the enormous body of the users and thus provides implications for industrial strategies such as targeted marketing. Here we propose a novel and statistically principled method, clust-LDA, which incorporates authorship structure into the topical modeling, thus accomplishing the task of the topical inferences across documents on the basis of authorship and, simultaneously, the identification of groupings between authors. We develop an inference procedure for clust-LDA and demonstrate its performance on simulated data, showing that clust-LDA out-performs the ``vanilla'' LDA on the topic identification task where authors exhibit distinctive topical preference.  We also showcase the empirical performance of clust-LDA based on a real-world social media dataset from Reddit.
\end{abstract}

\section{Introduction}

Targeted marketing has become one of the most prevalent strategies among consumer brands due to the efficiency of reaching audiences on digital media. With exponentiating numbers of online users engaging in social media and social network sites such as Twitter\footnote{\url{https://twitter.com}} and Reddit\footnote{\url{https://www.reddit.com}}, and the enormous volumes of content contributed by them in real time about subjects ranging from breaking news to trends in entertainment, social media corpora possess abundant information to inform corporate strategy and marketing. Efficient and effective text mining from these sources allows us to augment human interpretation with mathematical representations of the latent structure of the data, enabling us to gauge psychographic and behavioral tendencies of the underlying population, and apply targeting strategies, such as personalized advertising or niche marketing.

Probabilistic topic models such as Latent Dirichlet Allocation (LDA) are commonly used for analyzing the thematic structure of traditional large text corpora \cite{blei_2012}. 
However, social media corpora can pose distinctive challenges for topic models. The often-short length of the documents and the informal adherence to standard lexica induce high sparsity in the observed word co-occurrence matrices. Meanwhile, some social networks provide rich ancillary information about each document such as the characteristics of and the relationships between their authors. 

Much research has been done to improve the success of topic models on social media and similar corpora by incorporating additional meta-data about documents, and relationships between documents, into the model structure. The Author-topic model, in particular, creates structure at the author level by associating each author in the corpus with their own topic distribution \cite{rosen2004author}. Such structure, pooling across the high-resolution document level to support inference at the low-resolution author level, is particularly useful in support of targeted advertising to social media users. Nevertheless, social media corpora often have few documents per author to support accurate inference at the user level \cite{volkova2014inferring}. The motivates inference at even lower resolution, such as the grouping of authors.

In this paper, we propose a new topic model, clust-LDA, which groups authors into clusters with differentiated topic distributions. This approach simultaneously addresses the following motivations:
\begin{enumerate}
\item Jointly and coherently model both the underlying text structure and author group structure of the corpus;
\item Aggregate information hierarchically, yielding representations of the corpus at multi-resolutions from document, to author, and to group.
\item Pool information on the basis of authorship and author similarity to aid in the topical inference on short-length documents.
\end{enumerate}

In particular, the first of these motivations facilitates targeted marketing strategies based on audience segmentation.  Segmented strategies differentiate messaging and bidding at the group-level based on clustering in user behavior.  By directly modeling user group-level structure in textual discussions, topic model inferences can be used for marketing segmentation.

In the following article, we review related work, introduce the clust-LDA model, explore its theoretical performance using simulated data, and finally demonstrate its application to real world data collected from Reddit.

\section{Related Work}
\noindent\textbf{LDA.} Probabilistic models have been widely applied in the field of semantic analysis ~\cite{hofmann1999probabilistic,blei2003latent,sun2012probabilistic}.
In particular, LDA \cite{blei2003latent} has gained great popularity due to its interpretability and flexibility. LDA assumes that each document in a corpus is structured by a unique, underlying mixture of topics, where the topics are characterized by distinct distributions over the words in the vocabulary. For each word in a document, its topic is determined by the document-specific mixture that comes from a Dirichlet distribution, and subsequently the word is sampled from the topic-specific word distribution.

The original formulation of LDA (``vanilla"  LDA) has been broadly extended for a variety of tasks in text analysis. Blei et al.~\shortcite{griffiths2004hierarchical} generalized the topic structure of LDA into a hierarchy based on the nested Chinese restaurant process. Blei and McAuliffe~\shortcite{mcauliffe2008supervised} introduced supervised LDA to incorporate the value of a response variable for each document. Blei and Lafferty~\shortcite{blei2007correlated} imposed correlation among topics via the logistic normal prior in Correlated Topic Model (CTM).   Roberts et al.~\shortcite{roberts2013structural} further extended CTM to yield the STM, which incorporated document-level covariates and has yielded a wide range of flexibility and applications~\cite{roberts2014stm,roberts2014structural}.

\noindent\textbf{Author, community and clustering analysis.} Inference on the authorship and the communities or clusters within the authorship of corpora has also attracted great interests. Rosen-Zvi et al.~\shortcite{rosen2004author} first proposed the Author Topic (AT) model to incorporate the authorship of the documents into vanilla LDA by establishing multinomial distribution over topics for each author. Seroussi et al.~\shortcite{seroussi2012authorship} extended AT to model the author-specific topic distribution disjointly. These methods and others incorporate author-level information about documents, but do not consider the underlying structure among authors. 

A number of methods have been proposed to study the community structure among authors. Liu et al.~\shortcite{liu2009topic} combined an observed community structure of authors with LDA through a topic-link formulation.  Similarly, Lim et al.~\shortcite{lim2016nonparametric} explicitly modeled follower networks alongside a Dirichlet process formulation of text documents and McCallum et al.~\shortcite{mccallum2005author} modeled topics alongside sender-recipient links.  Li et al.~\shortcite{li2011modeling} and Yin et al.~\shortcite{yin2012latent} further integrated community modeling and LDA to facilitate latent community detection.  

Few past works have aimed at unsupervised clustering or grouping of authors to aggregate high-resolution information (individual, short documents) based on topic models to enable more accurate topical inference. Most existing methods\cite{jin2011transferring, xie2013integrating,yin2014dirichlet, zuo2016topic}  focused on clustering at the document level without considering authorship. Karandikar~\shortcite{karandikar2010clustering} explored techniques for clustering documents and authors in social media data independently, without seeking a joint model.

\section{Method}\label{sec:method}
The clust-LDA model aims at aggregating information across documents generated by the same author, and inferring a shared characterization of related authors based on the similarity in their documents.  We extend the AT model by introducing an additional hierarchy, a latent author cluster/group, and perform inference on a joint model for author clustering and topic modeling.
\begin{figure*}[!h]
\centering
\includegraphics[width=0.7\linewidth]{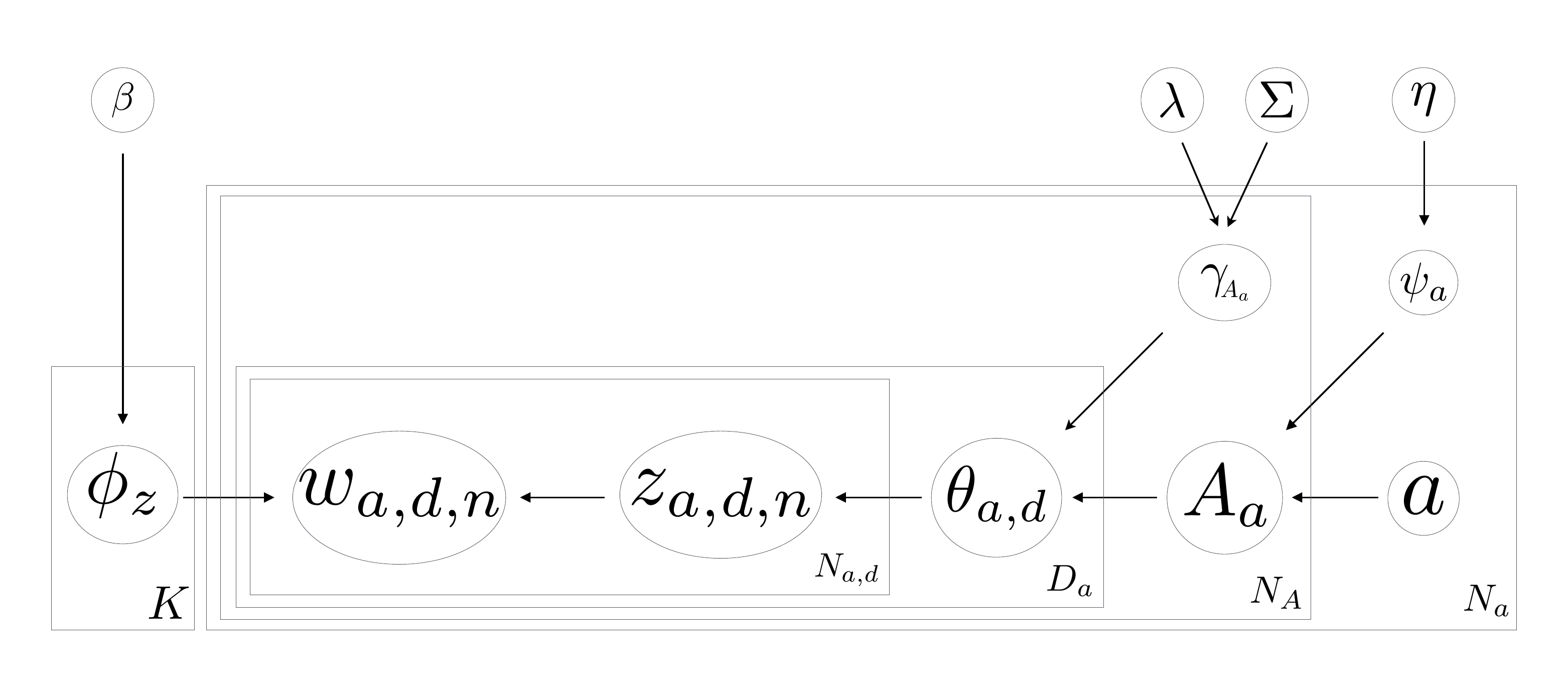}
\caption{Graphical representation of the clust-LDA model.}
\label{fig:graphic}
\end{figure*}

\subsection{Generative model}
We provide a graphical representation of clust-LDA in Figure \ref{fig:graphic}. The generative process is as follows:
\begin{enumerate}
\item For topic $k\in\{1,\dots, K\}$ and vocabulary size $V$,
\begin{enumerate}
\item Draw word distribution $\phi_k\sim {\rm Dirichlet}(\beta)$, where $\phi_k$ is $V$-length vector;
%
%
\end{enumerate}
\item For each group $A\in\{1, \dots, N_A\}$,
\begin{enumerate}
\item Draw cluster-level topic weight $\gamma_A\sim {\rm N_{K}}(0,\sigma_0^2I_K)$;
\end{enumerate}
\item For each author $a\in\{1,\dots, N_a\}$,
\begin{enumerate}
\item Draw group assignment distribution $\psi_a\sim \rm{Dirichlet}(\eta)$;
\item Draw the latent group membership $A_a\sim \rm{Multinom}(\psi_a)$;
\end{enumerate}
\item For each document $d\in \{1,\dots, D\}$,
\begin{enumerate}
\item Draw the document-level topic distribution $\theta_{a,d}|A_a, \gamma_{A_a}\sim \rm{logNormal}(\gamma_{A_a},\Sigma)$;
\item For each token $n \in \{1,\dots, N_{d}\}$,
\begin{enumerate}
\item Draw the topic $z_{a,d,n}\sim {\rm Multinom}(\theta_{a,d})$;
\item Draw the word  $w_{a,d,n}|(z_{a,d,n}=k)\sim \rm{Multinom}(\phi_k)$.
\end{enumerate}
\end{enumerate}
\end{enumerate}
The capitalized letters $\Phi$, $\Theta$, $\Gamma$, $\Psi$,  $Z$, $W$ denote the collections of their lower-case parameter/observation counterparts, i.e. $(\phi_k)$, $(\theta_{a,d})$, $(\gamma_{A_a})$, $(\psi_a)$, $(z_{a,d,n})$, $(w_{a,d,n})$ respectively.

\subsection{Inference: Expectation Conditional Maximization Algorithm}
We adopt a Gibbs-style iterative algorithm (or expectation conditional maximization, ECM) to infer the latent group label $A$ given the topic structure, i.e., the Cluster step (C-step),  and the topic model given the group membership, i.e. the Topic step (T-step).
\subsubsection{Cluster step}
Conditional on the parameters and variables from the topic model ($\Phi, \Theta, \Gamma, Z$), the posterior
\begin{align*}
p(A, \Psi|\Phi, \Theta, \Gamma,Z, W)\propto p(\Theta|\Gamma,A) p(A|\Psi) p(\Psi),
\end{align*}
which is equivalent to a Gaussian mixture model (GMM) with latent group label $A$ and observations $\Theta$ such that
\begin{align*}
\log(\theta_{a,d})|A_a, \gamma_{A_a}&\sim {\rm N}(\gamma_{A_a}, \Sigma)\\
A_a|\psi_{a}&\sim {\rm Multinom}(\psi_a)\\
\psi_a&\sim {\rm Dirichlet}(\eta).
\end{align*}
The expectation maximization (EM) algorithm provides the estimation for the posterior mode in the C-step (see \cite{benaglia2009mixtools}).

\subsubsection{Topic step}
Conditional on the group labels,
\begin{align*}
p(\Phi, \Theta, \Gamma, Z|W, \Psi, A)&\propto p(\Phi)p(\Gamma)\\
&p(\Theta|\Gamma, A)p(Z|\Theta)p(W|Z,\Phi),
\end{align*}
which is equivalent to the STM under a simplified configuration, in particular, an STM with each document's prevalence covariate set to its author's categorical group label, and no topical content modeling. Specifically, the generative process for the equivalent STM is:
\begin{enumerate}
\item For topic $k\in\{1,\dots, K\}$,
\begin{enumerate}
\item Draw word distribution $\phi_k\sim {\rm Dirichlet}(\beta)$;
\end{enumerate}
\item For each document $d\in \{1,\dots, D\}$ with its author being $a$, and author group being $A_a$,
\begin{enumerate}
\item Denote the document covariate $X_d=(x_{d,1},\dots,x_{d,N_A})^T$, where $x_{d,A}=1$ if and only if $A=A_a$;
\item Draw the document-level topic distribution $\theta_{a,d}|X_d, \Gamma \sim {\rm LN}(\Gamma X_d, \Sigma)$;
\item For each token $n \in \{1,\dots, N_{d}\}$,
\begin{enumerate}
\item Draw the topic $z_{a,d,n}\sim {\rm Multinom}(\theta_{a,d})$;
\item Draw the word  $w_{a,d,n}|z_{a,d,n}=k\sim {\rm Multinom}(\phi_k)$.
\end{enumerate}
\end{enumerate}
\end{enumerate}
The conditional STM can be readily inferred through variational methods based on a partially-collapsed EM as in Roberts et al.~\shortcite{roberts2014stm}.
\subsection{Diagnostics}
Here we discuss several diagnostics for model convergence and address the issue of multi-modality in the joint model.
\subsubsection{Convergence}
One of the major concerns with LDA-based topic model is the non-identifiability or weak identifiability of the model \cite{roberts2016navigating}.  In the case of clust-LDA, the label-switching issue in inferring author groups adds additional layer of complexity. The large dimensionality of model parameters in the LDA component renders it difficult to apply to clust-LDA traditional diagnostics for algorithm convergence, such as L2-difference of estimated parameters in consecutive steps.  As one of our major goals is to cluster the authors into groups, we compare the Rand index \cite{rand1971objective}, a label-free cluster similarity measure, between successive iterations as one diagnostic of convergence.

Define the convergence diagnostic at step $t$ as 
$
D_{c}^{(t)} = {\rm RI}(A^{(t-1)}, A^{(t)}),
$
where ${\rm RI}$ is the Rand index and $A^{(t)}$ is the inferred cluster labels at step $t$. We halt the iteration between C-step and T-step when $D_{c}^{(t)} > 1- \epsilon$. In practice, we can take $\epsilon=0$ for stability in author clustering.

\subsubsection{Multi-modality}
The iterative optimization and EM inference algorithm for the posterior mode only guarantees discovery and exploration around a local optimum, rather than the global optimum. One common approach for the global mode search is to perform multiple initializations. However, with the strong multi-modality and non-identifiability issues of LDA-based models, the variability of the converged models is high and, while a variety of options have been proposed \cite{wallach2009evaluation,mimno2011bayesian}, a widely accepted quantitative method to evaluate model fitness is lacking. In practice, human-driven qualitative investigation of the inferred topics and corresponding keywords is common best practice \cite{chang2009reading}, despite obvious drawbacks of subjectivity and inefficiency. Here we propose two quantitative diagnostics to evaluate fitness among the converged models from multiple initializations. 

For the $m$-th initialization, suppose 
\begin{enumerate}
\item The fitted logliklihood $D_{m}^{lik} = l(\hat{\Theta}; W,a)$, where $l$ is the logliklihood function of the joint model, and the $\hat\Theta$ is the estimates for all parameters. We choose the best fitted model from $m^{lik}= argmax_m D_{m}^{lik}$;
\item The clustering dispersion $D_{m}^{dips} = \frac{SSW}{SSB}=\frac{\sum_{j=1}^D\|\hat\theta_{a,d}-\exp\{\hat\gamma_{A_a}\}\|^2/D}{\sum_{i=1}^{N_A}\|\exp\{\hat\gamma_{i}\}-{\bar\exp\{\hat{\gamma}\}}\|^2/N_A}$, which measures the ratio between the within-cluster distance and between-cluster distance of data points. We choose the best fitted model from $m^{dips}= argmin_m D_{m}^{dips}$;
\end{enumerate}

\section{Experimental Results}
\subsection{Simulated Data}
To evaluate the performance of our model and algorithm, we simulate synthetic datasets from the generative process of clust-LDA.  We set $K=5$ topics, $N_A=3$ author groups; for hyper parameters, we set $\beta=1$ and vary $\eta$ and $\sigma_0^2$. Note that larger values of $\eta$ lead to less variance in $\psi_a$, i.e., greater similarity in topics, while larger values of $\sigma_0^2$ indicate stronger document-level noise.  For each dataset, we simulate $D = 852$ documents from $N_a=489$ authors, on average $N_{d}=20$. The scale of the synthetic dataset is motivated by the size of a representative thread from the Reddit network.

We adopt the iterative algorithm described in Section~\ref{sec:method}, and the Rand index-based diagnostic $D_{c}$ for convergence. For each combination of hyper-parameter settings $(\eta,\sigma_0^2)$, we simulate $n = 20$ datasets and, for each dataset, $M=20$ initializations. We choose the best fitted model based on the fitted log-likelihood $D^{lik}$. 

\begin{figure}[!h]
\centering
\includegraphics[width=\linewidth, page=1]{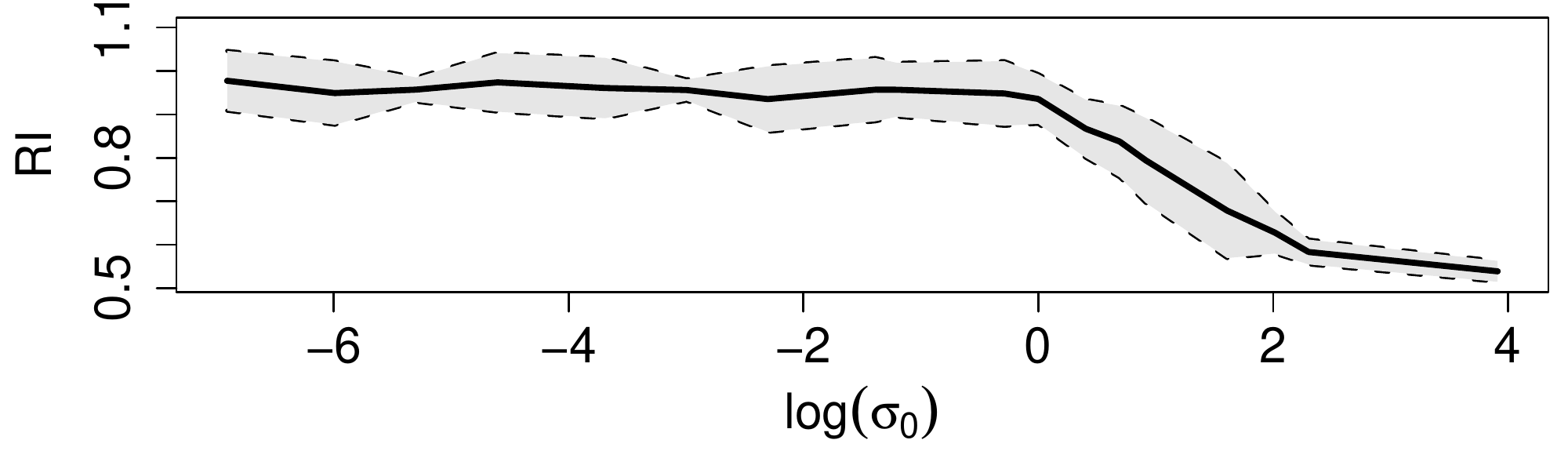}
\includegraphics[width=\linewidth,page=2]{plot2_lik_ri_sim_rep.pdf}
\caption{Efficiency of clust-LDA in recovering true author group assignments as a function of the hyperparameters $\sigma_0$ (top) and $\eta$ (bottom), measured by the Rand index and averaged across simulated datasets. }
\label{fig:RI}
\end{figure}
%
%

First, we report the clustering accuracy of clust-LDA based on the Rand index between the author group membership inferred by clust-LDA and the true simulated membership in Figure \ref{fig:RI}. In general, clust-LDA is able to provide accurate clustering with average Rand index ${\rm RI} > 0.9$ when the noise level and cluster topic similarity level are mild ($\eta<1$, $\sigma_0<1$). As the noise increases or the cluster variance decreases, the topics become more similar, raising the difficulty for inference. The average ${\rm RI}$ therefore decreases in these circumstances, as demonstrated in the figure. Notably, the variance in the clustering accuracy across the simulated datasets is rather consistent, which indicates the robustness and stability of the clustering method. 

\begin{figure}[!h]
\centering
\includegraphics[width=0.9\linewidth,page=1]{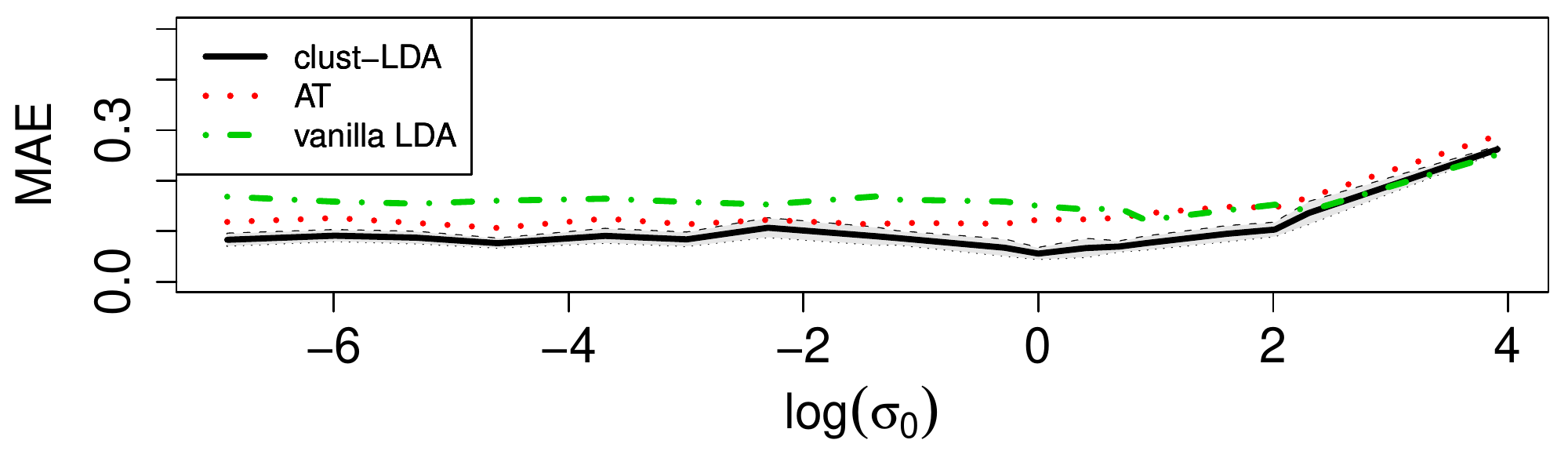}
\includegraphics[width=0.9\linewidth,page=2]{{plot2_max_ri_gamma_mae_auth1_sim_rep}.pdf}
\caption{The average MAE of the author-group level topic distribution ($\gamma$) on author level across simulated datasets, for varying values of the hyperparameters $\sigma_0$ (top) and $\eta$ (bottom), comparing clust-LDA with vanilla LDA and AT.}
\label{fig:gamma}
\end{figure}

We also evaluate the estimation accuracy of clust-LDA on the cluster-level topic distribution parameter $\Gamma=(\gamma_1,\gamma_2,\gamma_3)$ as measured by the mean absolute error (MAE).  We compare the performance of clust-LDA with the vanilla LDA and AT models in Figure \ref{fig:gamma}. For a fair comparison with these two benchmark methods, the MAE is defined as 
$
MAE = \frac{1}{N_a}\sum_{a}\|\hat{\gamma}_{A_a}-\gamma_{A_a}\|_1,
$
where, for clust-LDA, $\hat{\gamma}_{A_a}$ is the estimated topic distribution of the inferred group of author $a$; for AT, $\hat{\gamma}_{A_a}$ is the estimated author-level topic distribution; and for vanilla LDA, $\hat{\gamma}_{A_a}$ is estimated by the average of the document-level topic distribution $\hat{\theta}_d$ across the documents of author $a$. Note that all three methods are compared at the level of the topic distribution for each author for fairness and no extraneous information (e.g. the true cluster labels) is included for the MAE evaluation of any method.  Both the AT and vanilla LDA are implemented by the \textit{stm} \textit{R}~package \cite{roberts2014stm}.

In general, clust-LDA shows strong advantages over two benchmark methods in terms of the MAE, especially when the noise level or the cluster similarity are low to moderate ($\sigma_0 < 10$, or $\eta < 2$).   As the noise level increases ($\sigma_0 > 10$), LDA gradually converges to the performance of clust-LDA as the marginal value of information provided by author clustering decreases. Such observation gives us confidence in clust-LDA for improved topical inference on corpora with strong and distinctive author grouping characterization.

\subsection{Reddit Data}
We confront clust-LDA with a real world sample dataset from the Reddit network. To test the performance of our method, we combine the comments of three Reddit forums (``subreddits'') of television shows, namely {\it Evangelion} (\url{https://www.reddit.com/r/evangelion/}), {\it It's Always Sunny in Philadelphia} (\url{https://www.reddit.com/r/IASIP/}), and {\it The Simpsons} (\url{https://www.reddit.com/r/TheSimpsons/}). The integrated corpus contains 6,997 comments (documents) from 1,662 Reddit users (authors) with a vocabulary of 1,093 words, after removing stop words and low-frequency words (less than $10$ counts in the corpus). We run clust-LDA on the integrated corpus (with all three shows) with the number of topics $K=20$, number of groups $N_A=3$, and $M=20$ initializations.
\begin{table}[!h]
\centering\caption{Contingency table of inferred cluster and observed subreddit membership by clust-LDA, in percentages ($\%$)}\label{tab:clust}
\small
\begin{tabular}{rrrr}
  \hline
Subthread & Cluster I &Cluster II & Cluster III \\ 
     \hline
 Evangelion& 7.56 & 6.19 & 87.22 \\ 
Sunny & 42.25 & 21.48 & 41.03 \\ 
Simpsons & 53.37 & 25.69 & 18.38 \\ 
   \hline
\end{tabular}
\end{table}

\begin{table*}[!h]
\centering\caption{Top words per topic, topic distributions per cluster, and TV show associations inferred by clust-LDA (labeled when strongly identifiable).}\label{tab:topwords}
\tiny
\begin{tabular}{c|lllll|lll|l}
  \hline
{Topic} & \multicolumn{5}{c}{{Top words}} &  {Cluster} I & {Cluster} II & {Cluster} III &{TV show} \\ 
  \hline
1 & want & well & right & shit & last &  \textbf{0.051} & \textbf{0.054} & 0.040 & -- \\ 
  2 & eva & use & seri & anim & evangelion & 0.025 & 0.021 & \textbf{0.051} & Evangelion \\ 
  3 & like & look & lot & find & mani & \textbf{0.057} & \textbf{0.061} & 0.049 & -- \\ 
  4 & peopl & charact & even & way & still & 0.049 & 0.042 & \textbf{0.050} & -- \\ 
  5 & get & think & never & mean & rememb & \textbf{0.053} & \textbf{0.051} & 0.049 & -- \\ 
  6 & pretti & work & back & sure & yes & 0.049 & \textbf{0.050} & 0.047 & -- \\ 
  7 & just & year & mayb & let & guess & 0.047 & 0.048 & 0.046 & -- \\ 
  8 & mcdonald & god & bitch & buy & car & 0.012 & 0.020 & 0.009 & Sunny \\ 
  9 & asuka & rei & fuck & nge & unit & 0.023 & 0.024 & \textbf{0.055} &  Evangelion\\ 
  10 & realli & feel & alway & new & sinc & 0.043 & 0.035 & 0.039 & -- \\ 
  11 & time & show & got & favorit & tell & 0.048 & 0.041 & 0.039 & -- \\ 
  12 & also & actual & said & around & bad & 0.040 & 0.045 & 0.036 & -- \\ 
  13 & one & watch & will & first & two & 0.058 & 0.046 & 0.048 &  -- \\ 
  14 & make & can & thing & much & yeah & \textbf{0.051} & \textbf{0.052} & 0.046 &  -- \\ 
  15 & know & now & probabl & play & done & 0.049 & 0.044 & 0.047 & -- \\ 
  16 & mac & charli & denni & frank & man & \textbf{0.081} & \textbf{0.109} & 0.048 & Sunny \\ 
  17 & shinji & end & need & seem & happen & 0.027 & 0.025 & \textbf{0.057} & Evangelion \\ 
  18 & love & see & good & come & scene & \textbf{0.063} & \textbf{0.066} & 0.046 & -- \\ 
  19 & say & someth & thank & movi & reason & \textbf{0.052} & \textbf{0.057} & 0.044 & -- \\ 
  20 & episod & season & best & simpson & better & \textbf{0.079} & \textbf{0.062} & 0.043 & Simpsons \\ 
   \hline
\end{tabular}
\end{table*}

\begin{figure}[!h]
\centering
\includegraphics[width=0.9\linewidth]{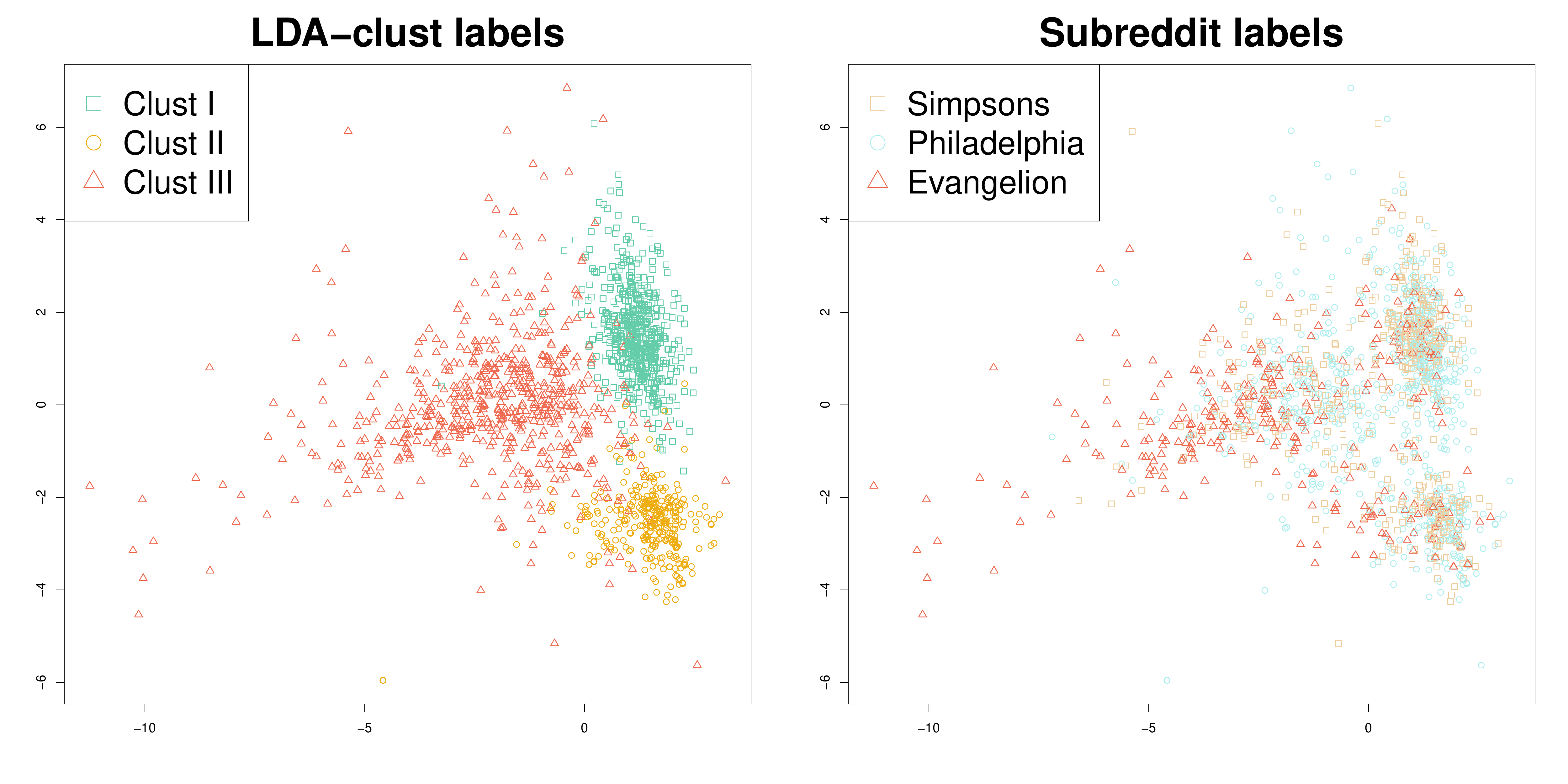}
\caption{Author-cluster illustration. Scatter plot of author-level topic distributions based on the first two principle components. Different colors and shapes indicate different clust-LDA inferred clusters (left) and original subreddit labels (right).}
\label{fig:author-scatter}
\end{figure}

We report the results of the inferred topic model and cluster features in Tables \ref{tab:topwords} - \ref{tab:clust}  and Figure \ref{fig:author-scatter}. If we take the subreddit identity to be the ground truth of an author group, the Rand index between the assumed group labels and our inferences is ${\rm RI} = 0.63$ for the model with the minimum dispersion ($D_m$). Table \ref{tab:clust} summarizes the correspondence between the inferred clusters and the original subreddits for all the documents, while Figure~\ref{fig:author-scatter} illustrates this relationship.  Cluster~III corresponds closely to the subreddit of {\it Evangelion}, indicating that most comments about {\it Evangelion} are strongly clustered together under the clust-LDA model. This observation is also reflected in Table~\ref{tab:topwords}, which shows that the top weighted words among the top topics of Cluster~III include a great number of keywords specific to {\it Evangelion}, such as {\it shinji, gendo, kaworu, asuka, {\rm and} evangelion}. In contrast, Clusters~I \& II are shown to be nearly even mixtures of content from the other two subthreads. The top weighted topics of Clusters~I \& II overlap heavily.  While the clusters are mixed, some of the topics are strongly associated with individual shows.  Topic~20 and Topic~16 contain the keywords specific to {\it The Simpsons} and 
{\it It's Always Sunny in Philadelphia}, respectively (Table \ref{tab:topwords}). Such results may attribute to the varying identifiability of the three subreddits: {\it Evangelion} is a Japanese cartoon series that contains lots of distinctive keywords in Japanese, while the other two are American TV comedies with more commonality in discussion keywords, themes, and topics.
\section{Discussion}
Throughout the experiments on clust-LDA, we assumed the number of cluster $N_A$ is known and fixed. Yet in real application, the determination of cluster number $N_A$ is a generic challenge for unsupervised learning tasks and has been extensively studied. One suggestion in practice is to adopt established diagnostics such as AIC, BIC, silhouette or gap statistics \cite{kodinariya2013review, rousseeuw1987silhouettes, tibshirani2001estimating} to determine $N_A$ in the C-step. Detailed discussion is provided in the Supplement.
\section{Conclusion}
In this article, we present clust-LDA, a joint model for inference on both the topic structure and the author group/cluster of text datasets. Clust-LDA introduces a latent author group/cluster hierarchy to the traditional LDA model to discover and characterize groups of authors in terms of their topical preferences. Through experimental studies on synthetic data and empirical samples of Reddit social network data, we have shown that clust-LDA is effective in extracting author groupings and the topical preferences of authors, particularly when author groups exhibit well-separated topical preferences and the noise level is relatively low. This indicates that clust-LDA is most useful for short-length, non-prolific (low number of documents per author) copora where the similarity among texts is best characterized by grouping of authors. 

With the ever-growing abundance of short-length social media data, clust-LDA can serve as a tool for text mining efforts that focus on author characterization and can particularly aid segmentation-based targeted marketing strategies.

\bibliography{clust_LDA}
\bibliographystyle{acl_natbib}

\end{document}